# Pressure-induced volumetric negative thermal expansion in CoZr$_2$ superconductor

*Yuto Watanabe*[A], *Hiroto Arima*[A], *Saori Kawaguchi-Imada*[B], *Hirokazu Kadobayashi*[B], *Kenta Oka*[B], *Hidetomo Usui*[C], *Ryo Matsumoto*[D], *Yoshihiko Takano*[D], *Takeshi Kawahata*[E], *Chizuru Kawashima*[E], *Hiroki Takahashi*[E], *Aichi Yamashita*[A], *Yoshikazu Mizuguchi*[A*]

A. Department of Physics, Tokyo Metropolitan University, 1-1, Minami–Osawa, Hachioji 192-0397, Japan.
B. Japan Synchrotron Radiation Research Institute (JASRI), 1-1-1 Koto, Sayo-cho, Sayo-gun Hyogo 679-5198, Japan.
C. Department of Applied Physics, Shimane University, Matsue, Shimane 690-8504, Japan
D. International Center for Materials Nanoarchitectonics (MANA), National Institute for Materials Science, Tsukuba, Ibaraki 305-0047, Japan.
E. Department of Physics, College of Humanities and Sciences, Nihon University, Setagaya, Tokyo 1568550, Japan.

E-mail: mizugu@tmu.ac.jp

Keywords: superconductor, pressure, negative volumetric thermal expansion

We investigate the thermal expansion and superconducting properties of a CuAl$_2$-type (tetragonal) superconductor CoZr$_2$ under high pressures. We perform high-pressure synchrotron X-ray diffraction in a pressure range of 2.9 GPa < $P$ < 10.4 GPa and discover that CoZr$_2$ exhibits volumetric negative thermal expansion under high pressures. Although the uniaxial positive thermal expansion (PTE) along the *a*-axis is observed under ambient pressure, that is suppressed by pressure, while the large uniaxial negative thermal expansion (NTE) along the *c*-axis is maintained under the pressure regime. As a result of a combination of the suppressed uniaxial PTE along the *a*-axis and uniaxial NTE along the *c*-axis, volumetric negative thermal expansion is achieved under high pressure in CoZr$_2$. The mechanisms of volumetric NTE would be based on the flexible crystal structure caused by the soft Co-Co bond as seen in the iso-structural compound FeZr$_2$, which exhibits uniaxial NTE along the *c*-axis. We also perform high-pressure electrical resistance measurements of CoZr$_2$ to confirm the presence of superconductivity under the examined pressure regime in the range of 0.03 GPa < $P$ < 41.9 GPa. We confirm the presence of superconductivity under all pressures and observe dome-like shape pressure dependence of superconducting transition temperature. Because of the coexistence of two phenomena, which are volumetric NTE and superconductivity, in CoZr$_2$ under high pressure, the coexistence would be achievable under ambient pressure by tuning chemical compositions after our present observation.

**1. Introduction**

Thermal expansion is one of the most significant characteristics of materials because it is connected to the crystal structure and electronic structure, which determines the physical properties of materials. In most cases, materials expand upon heating; this conventional property is called positive thermal expansion (PTE). In contrast, negative thermal expansion (NTE) is defined as contraction upon heating, and such NTE has been observed in various materials.[1,2,3,4] The mechanisms of NTE are diverse and correlated to the flexible crystal structure,[5,6] phase transition,[7] magnetic order-to-disorder transition,[8] and/or so on. NTE has been used to achieve zero thermal expansion in practical devices by making a composite using PTE and NTE materials. In superconducting devices, the heat cycle between working (very low) temperature and room temperature when turning off the device is a critical issue because the heat cycle degrades the surface and junction between different materials. If a superconductor with NTE is in a wide temperature range below room temperature, the heat-cycle problem will be improved. Isotropic and uniaxial NTE has been reported in various superconducting materials, such as single elements Nb[9,10] and Ta,[10] layered materials MgB$_2$,[9,11] YBa$_2$Cu$_3$O$_{7-\delta}$,[12] Bi$_2$Sr$_2$CaCu$_2$O$_{8+x}$,[13] PrFeAsO,[14] and Ba(Fe$_{1-x}$Co$_x$)$_2$As$_2$ ($x$ = 0.16, 0.23).[15] Those NTEs have been, however, observed in a limited temperature range, and volumetric NTE in a wide temperature range has not been achieved in any bulk superconducting materials.

Recently, we observed uniaxial NTE with a wide temperature range in CuAl$_2$-type (tetragonal) $Tr$Zr$_2$ and $Tr$Zr$_3$ ($Tr$: transition metal), which are superconductors.[16,17,18] In the $Tr$Zr$_2$ system, we revealed that uniaxial NTE along the $c$-axis can be controlled by lattice-constant ratio, $c/a$, through chemical element substitution.[19,20] The anomalous bonding states related to uniaxial NTE along the $c$-axis in $Tr$Zr$_2$ have been observed using X-ray absorption spectroscopy as well.[21] Moreover, Xu *et al.* revealed that FeZr$_2$ exhibits giant uniaxial NTE along the $c$-axis, and they proposed that the soft Fe-Fe bond and flexible structure caused by optical phonons play an important role in the origin of uniaxial NTE along the $c$-axis.[22] Therefore, the crystal structural modification should be critical to the NTE phenomena in $Tr$Zr$_2$.

Herein, we show the observation of volumetric NTE in CoZr$_2$ under high pressure. At ambient pressure, CoZr$_2$ shows superconductivity at $T_c$ = 5.8 K ($T_c$: superconducting transition temperature), uniaxial PTE along the $a$-axis, and uniaxial NTE along the $c$-axis. The uniaxial

PTE along the *a*-axis is suppressed by pressure, but the uniaxial NTE along the *c*-axis was not suppressed by pressure. As a consequence of competition between uniaxial PTE and NTE along the *a*- and *c*-axes, the volumetric NTE is realized. Since the coexistence of superconductivity and volumetric NTE in a wide temperature range is quite rare, we confirmed the presence of superconductivity in CoZr$_2$ under high pressures by electrical resistance measurements.

## 2. Results and discussion
### 2.1. Thermal expansion under high pressure

We show the schematic images of the thermal expansions of a structural analogue NiZr$_2$ and CoZr$_2$ (under ambient pressure and high pressure) in Figure 1. These compounds have a tetragonal CuAl$_2$-type crystal structure (space group: *I*4/*mcm*). NiZr$_2$ shows PTE both along the *a*- and *c*-axes, and thus, the coefficient of volumetric thermal expansion $\beta$ is positive (Figure 1(a)).[19,20] In the tetragonal crystal structure, $\beta$ can be calculated to the following equation:

$$\beta = 2\alpha_a + \alpha_c \tag{1}$$

where $\alpha_a$ and $\alpha_c$ are the coefficients of linear thermal expansion along *a*- and *c*-axes, respectively. In contrast to NiZr$_2$, CoZr$_2$ exhibits uniaxial NTE along the *c*-axis (Figure 1(b)).[16] For CoZr$_2$, $\alpha_a$ and $\alpha_c$ at ambient pressure are $21 \pm 6$ μK$^{-1}$ and $-18 \pm 5$ μK$^{-1}$, respectively. Therefore, we obtain $\beta = 24 \pm 1$ μK$^{-1}$ (see Figure S1 in the Supporting information). As discussed later, under high pressures, we find that the uniaxial PTE along the *a*-axis is suppressed, but uniaxial NTE along the *c*-axis is not suppressed; as a consequence, $\beta$ could be a negative value, which means volumetric NTE is realized in CoZr$_2$ under pressures (Figure 1(c)).

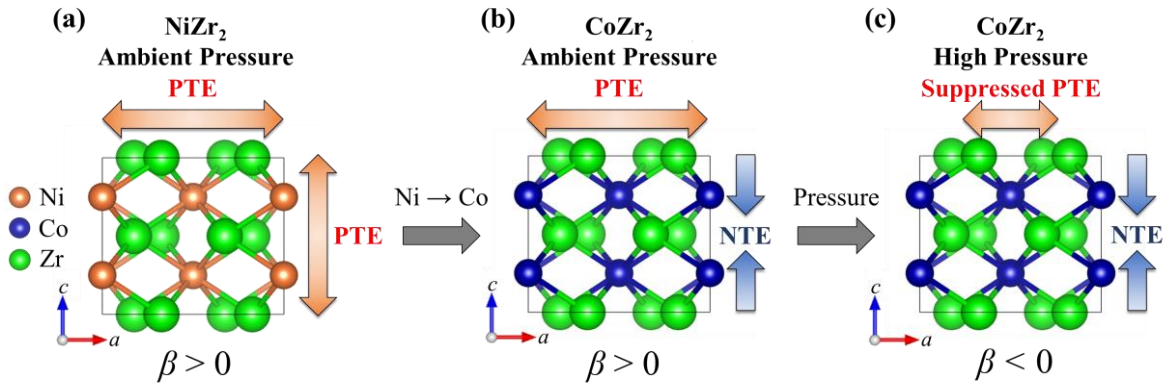

**Figure 1**. Schematic images of thermal expansion (a) NiZr$_2$ under ambient pressure, (b) CoZr$_2$ under ambient pressure, and (c) CoZr$_2$ under high pressure.

To investigate the properties of thermal expansion under high pressure, we performed high-pressure synchrotron X-ray diffraction (HP-SXRD) using a diamond anvil cell (DAC). Figure 2(a) shows the HP-SXRD patterns taken at $T = 303$ K under $P = 2.9, 5.0, 6.7, 7.4$, and $10.4$ GPa. The crystal structure remains tetragonal $CuAl_2$-type up to $P = 10.4$ GPa. We observe a shift of the 002 and 220 peaks toward the higher angle side by applying pressure as shown in Figures 2(b) and 2(c). The shifts of the peaks result in decreasing lattice constants $a$ and $c$. Figure 2(d) shows the pressure dependence of the lattice constants. Lattice constants normalized by a value at ambient pressure ($a_0$ and $c_0$) are shown in Figure 2(e). The $a$-axis is stiffer than the $c$-axis under high pressure. This implies that the crystal structure of $CoZr_2$ along the $c$-axis direction is more flexible to pressure. The same trend of $a$ and $c$ against pressure was also observed in laboratory experiments, which were performed using another DAC with a Mo-K$\alpha$ radiation (see Figure S2 in the Supporting information). As mentioned in the introduction part, $FeZr_2$, which is an iso-structural compound with $CoZr_2$ and $NiZr_2$, exhibits giant uniaxial NTE along the $c$-axis.[22] They revealed that the strong Fe$3dz^2$-Fe$3dz^2$ interaction can play an important role in stabilizing large $c/a$ in the $CuAl_2$-type crystal structure and contributes to the soft Fe-Fe bond, which provides a large contraction space along the $c$-axis. Furthermore, the optical phonons with a phonon energy of several meV make the flexible structure in $FeZr_2$, which leads to giant uniaxial NTE along the $c$-axis. These flexible characteristics of crystal structure would be common to $CoZr_2$ because it has the same crystal structure with a similar $c/a$ ratio as $FeZr_2$ and shows large uniaxial NTE along the $c$-axis. Figure 2(f) shows the pressure dependence of lattice volume $V$ of $CoZr_2$. The solid line is the fit to the 3rd-order Birch-Murnaghan formula as expressed following equation:[23]

$$P = \frac{3}{2}B_0 \left\{ \left(\frac{V_0}{V}\right)^{\frac{7}{3}} - \left(\frac{V_0}{V}\right)^{\frac{5}{3}} \right\} \left\{ 1 + \frac{3}{4}(B_0' - 4)\left[\left(\frac{V_0}{V}\right)^{\frac{2}{3}} - 1\right]\right\} \quad (2)$$

where $V_0$, $B_0$, and $B_0'$ are the volume at ambient pressure, bulk modulus, and 1st-order pressure derivative of $B_0$, respectively. As the fitting result, we obtain $B_0 = 100 \pm 6$ GPa and $B_0' = 6 \pm 2$. The obtained $B_0$ value is close to the computational calculation results, $B_0 = 134.81$[24] and 129 GPa[25]. We note here that the 3rd-order Birch-Murnaghan formula assumes the cubic structure, thus, the obtained $B_0$ value can deviate from the actual value.

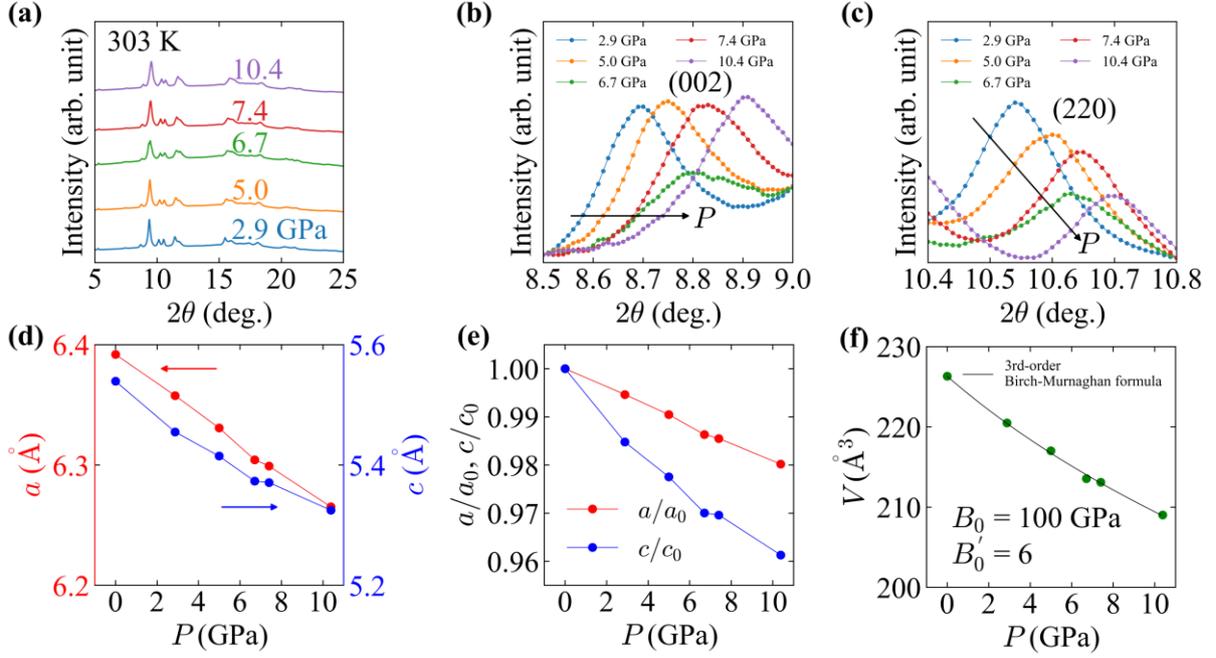

**Figure 2**. (a) HP-SXRD patterns at $T$ = 303 K for CoZr$_2$. (b,c) A shift of 002 and 220 peaks due to pressure. Pressure dependence of (d) lattice constants $a$ and $c$, (e) $a$ and $c$ normalized by a value at ambient pressure, (f) lattice volume $V$. The solid line is fit to the 3rd-order Birch-Murnaghan formula.

Next, we show the results of the thermal expansion of CoZr$_2$ at $P$ = 2.9 GPa as an example of the high-pressure data set. For all the experiments under pressures, we evaluated the fluctuation of applied pressures because the thermal expansion is easily affected by pressure changes. The data shown in this paper has been carefully taken with the manner. Figure 3(a) shows the HP-SXRD patterns at $P$ = 2.9 GPa under temperatures ranging from $T$ = 303 K to 453 K with an increment of 10 K. There was no crystal structural transition between the temperature region under $P$ = 2.9 GPa, which was commonly confirmed in other all applied pressures (see Figure S3 in the Supporting information). The absence of crystal structural transition at ambient pressure was confirmed in the wide temperature (7 K < $T$ < 572 K) in Ref. 16. As the temperature increased, a clear shift of the 002 peak toward the higher angle side is observed as shown in Figure 3(b). However, the 220 peak position is almost the same with pressure as shown in Figure 3(c). The robustness of the $a$-axis to pressure seen from the 220 peak results in a small value of $\alpha_a$ = 9 ± 1 μK$^{-1}$, which is clearly smaller than the value at ambient pressure (see Figure S1 in the Supporting information). Figures 3(d) and 3(e) are the temperature dependence of $a$ and $c$ at $P$ = 2.9 GPa. Even under pressure, the large $c$-axis NTE is present with $\alpha_c$ = -24 ± 1 μK$^{-1}$, which is almost the same as that observed under ambient pressure. From Equation (1), we obtain $\beta$ = -6 ± 2 μK$^{-1}$, suggesting the pressure-induced volumetric NTE in $CoZr_2$. Figure 3(f) shows the temperature dependence of $V$ at $P = 2.9$ GPa. We see the volume slightly contracted as the temperature increased.

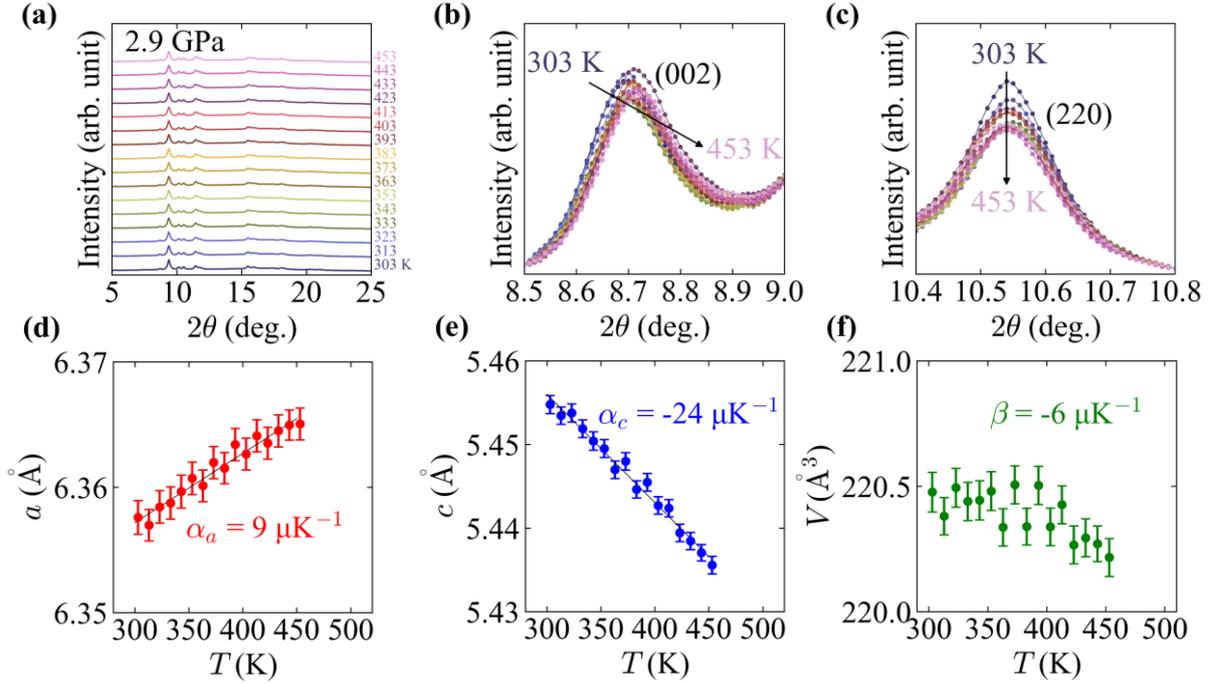

**Figure 3**. (a) HP-SXRD patterns at $P = 2.9$ GPa for $CoZr_2$. (b,c) A shift of 002 and 220 peaks due to heating. Temperature dependence of lattice constants (d) $a$, (e) $c$. The solid line is fit to the linear line. (f) lattice volume $V$.

We summarize the pressure dependence of $\alpha_a$, $\alpha_c$, and $\beta$ in Figure 4. When the pressure is applied to $CoZr_2$, the uniaxial PTE along the $a$-axis is suppressed; thus, $\alpha_a$ under high pressure is lower than at ambient pressure (Figure 4(a)). In contrast, even under high-pressure conditions, the uniaxial NTE along the $c$-axis is not suppressed; therefore, $\alpha_c$ is almost independent of pressure (Figure 4(b)). Above the ambient pressure, the $\beta$ value can be negative because the impact of uniaxial NTE along the $c$-axis on the volume exceeds the suppressed uniaxial PTE along the $a$-axis, resulting in volumetric NTE. The results of Rietveld refinement of $CoZr_2$ under $P = 2.9$ GPa (303 K, 453 K) and $P = 10.4$ GPa (303K, 403 K) are shown in Figure S4 (see Supporting information).

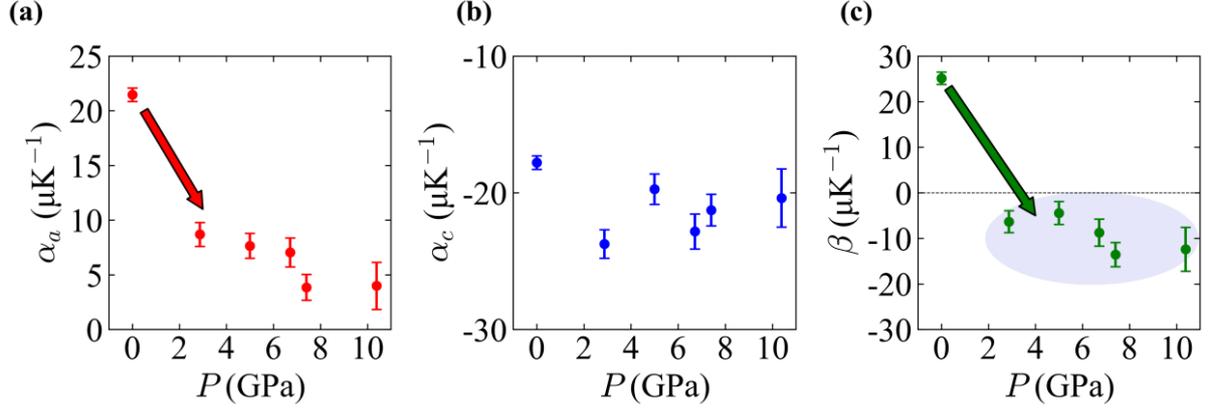

**Figure 4**. Pressure dependence of the coefficient of linear thermal expansion (a) along the $a$-axis $\alpha_a$, (b) along the $c$-axis $\alpha_c$. (c) Pressure dependence of the coefficient of volumetric thermal expansion $\beta$. In the tetragonal crystal structure, $\beta$ can be calculated to $\beta = 2\alpha_a + \alpha_c$.

2.2. Superconducting properties

We measured the electrical resistance ($R$) of CoZr$_2$ under high pressures (0.03 GPa $< P <$ 41.9 GPa) to confirm the presence of superconductivity. Figures 5(a) and 5(b) show the temperature dependence of $R$ under pressures. As temperature decreases, $R$ decreases with a negative curvature, which is a trend commonly seen in $d$-electron superconductors.[26, 27] At low temperatures, the $R$ drops to zero at $T_c$ under all applied pressures. The dome-shaped pressure dependence of $T_c$ is observed as shown in Figure 5(c). The $T_c$ taken from the $R$ data at ambient pressure ($P = 0.03$ GPa) is 5.8 K, consistent with the value taken from magnetic susceptibility measurement at ambient pressure (see Figure S5 in the Supporting information). As pressure increases, the $T_c$ increases up to $P = 17.9$ GPa and reaches 6.5 K, but the trend of $T_c$ changed at $P > 17.9$ GPa; $T_c$ decreases with pressure at higher pressures. In a study on a single crystal of CoZr$_2$, the $T_c$ reached 9.5 K at $P = 8$ GPa,[26] which is higher than the highest $T_c$ obtained in this study with polycrystalline CoZr$_2$. The discrepancy of the highest $T_c$ values may be due to the difference in the reactions of $T_c$ to generated pressure caused by the experimental conditions: pressure cells and sample type (single or poly crystals). In the low-temperature region where $T_c < T \ll \Theta_D$ (Debye temperature), the $R$ could be fitted to the power-law relation:

$$R(T) = R_0 + AT^n \tag{3}$$

where $R_0$ is the residual resistance, $A$ is the numerical temperature-independent coefficient, and $n$ is a component depending on the carrier scattering mechanisms. We used $R$ at 10 K $< T <$ 30 K under pressures in the fitting to power-law relation, which yielded $n \sim 3$ for all applied pressures as shown in Figure S6 (see Supporting information) The $T^3$ dependence on low-temperature $R$ can be explained with a phonon-assisted $s$-$d$ electron scattering model.[28] The $R$ of compounds composed of *d*-block elements is empirically known that it could be fitted to the Parallel-resistor model[27,29] developed by Wiesmann *et al.*[30] In the model, the $R$ is described as the following equation:

$$R(T) = \left[\frac{1}{R_{\text{sat}}(T)} + \frac{1}{R_{\text{ideal}}(T)}\right]^{-1} \quad (4)$$

where $R_{\text{sat}}$ is the saturated $R$ at high temperatures. Fisk and Webb found that, at high temperatures, the $R$ of strong-coupled superconducting transition-metal compounds such as $Nb_3Sn$ and $Nb_3Sb$ saturates at a certain value that corresponds to an electron mean free path of the order of the interatomic spacing in the compound.[31] $R_{\text{ideal}}$ is composed of residual electrical resistance $R_0$ and phonon-assisted *s-d* electron scattering term as the following expression:

$$R_{\text{ideal}}(T) = R_0 + C\left(\frac{T}{\Theta_{\text{D}}}\right)^3 \int_0^{\frac{\Theta_{\text{D}}}{T}} \frac{x^3}{(e^x - 1)(1 - e^{-x})} dx \quad (5)$$

where $C$ is the numerical temperature-independent coefficient. We used the $R$ data at 10 K $< T <$ 300 K for fitting to the Parallel-resistor model and evaluated the $\Theta_{\text{D}}$ values as a function of applied pressure as shown in Figure 5(d). The calculated $\Theta_{\text{D}}$ is 292 K at ambient pressure ($P =$ 0.03 GPa), which slightly deviates from the value $\Theta_{\text{D}} = 260$ K obtained by another experimental result using specific heat measurement.[32]

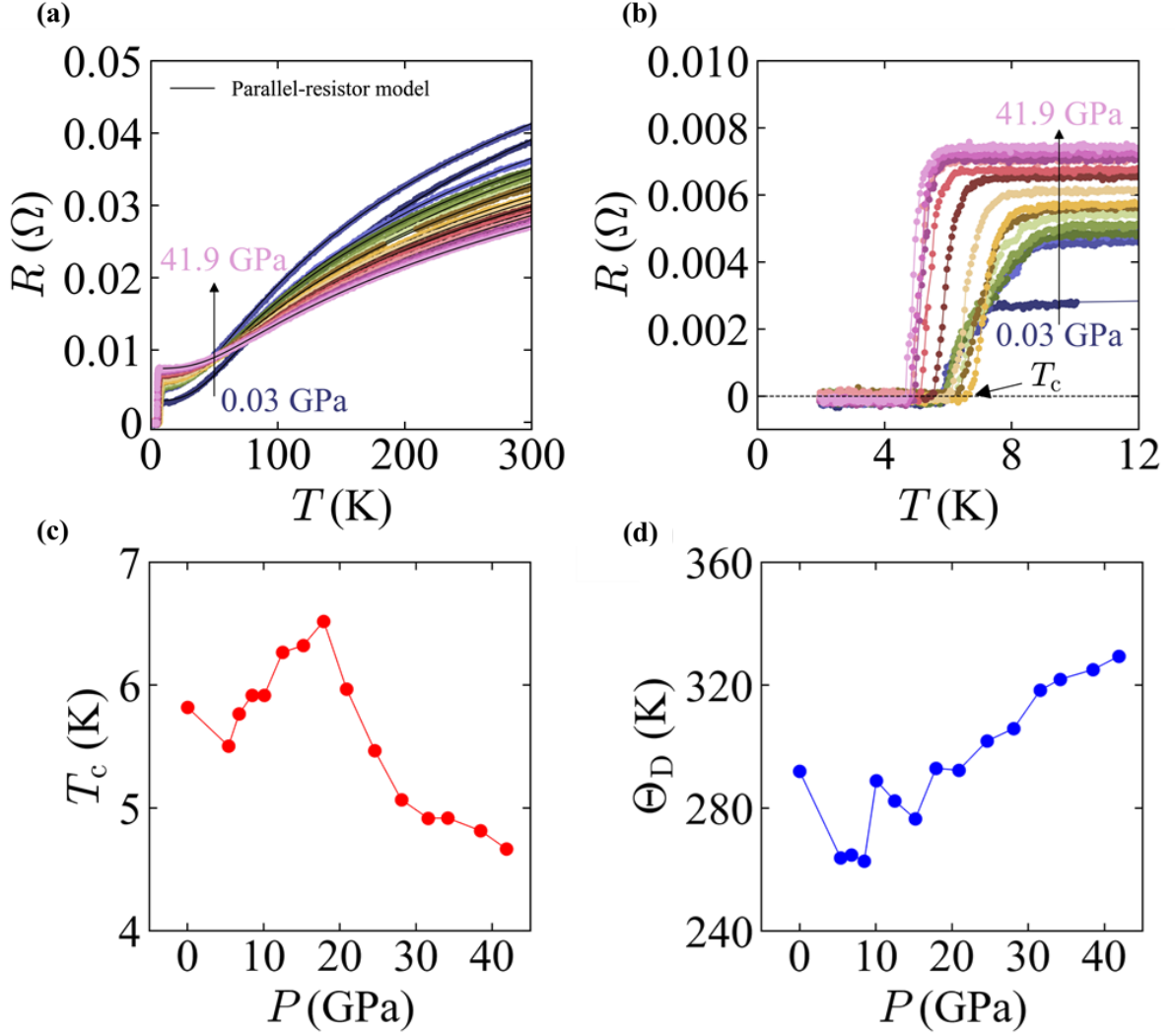

**Figure 5**. (a,b) Temperature dependence of electrical resistance ($R$) under applied pressures. The solid lines are fit to the Parallel-resistor model. (c,d) Pressure dependence of superconducting transition temperature $T_c$ and Debye temperature $\Theta_D$.

As mentioned above, we observed the dome-shaped pressure dependence of $T_c$ (Figure 5(c)). In conventional weak-coupling electron-phonon (BCS) superconductors, $T_c$ can be expressed by the following equation:[33]

$$T_c = 1.13\Theta_D \exp\left\{-\frac{1}{N(0)U}\right\} \tag{6}$$

where $N(0)$ is the electronic density of states at the Fermi energy, and $U$ is the effective Coulomb interaction constant. According to Equation (6), the $T_c$ is mostly controlled by $\Theta_D$ and $N(0)$. The $\Theta_D$ gradually increases upon applying pressure (Figure 5(d)), contributing to an enhancement of $T_c$. In contrast, $N(0)$ usually decreases under high-pressure conditions because of the expansion of bandwidth,[34] contributing to the suppression of $T_c$. Therefore, the competition of the contributions of $\Theta_D$ and $N(0)$ to pairing would cause the dome-shaped pressure dependence of $T_c$. The dome-shaped pressure dependence of $T_c$ is observed in other superconductors such as CaSb$_2$,[35] CeV$_3$Sb$_5$,[36] AuTe$_2$,[37] and Cd$_2$Re$_2$O$_7$[38]. The possible causes of causing discontinuation or dome shaped pressure dependence of $T_c$ are proposed as crystal structural transition or Lifshitz transitions.

$\Theta_D$ is related to the elastic properties of the material, especially the stiffness.[39] As well known, diamond or crystals with a diamond-type structure whose large $\Theta_D$ exhibits a small linear thermal expansion coefficient.[40,41] Therefore, the trend of $\Theta_D$ against applied pressure as shown in Figure 5(d) is consistent with the decrease of $\alpha_a$ under pressure. The pressure effect on $\alpha_c$ would be negligible because of the Co-Co soft bond like the Fe-Fe soft bond seen in FeZr$_2$.[22]

### 2.3 Electrical structure

Here, we show the total electronic density of states (DOS) and partial DOS in Figure 6. The value of the total DOS near the Fermi energy ($E_F$) is comparable to each other at $P = 0$, 5, and 10 GPa (Figure 6(a)). This trend suggests that the changes in $T_c$ under pressures is dominated mainly by $\Theta_D$ up to $P = 10$ GPa. The partial DOS for Co-3$d$ and Zr-4$d$ orbitals is shown in Figures 6(b) and 6(c) at $P = 0$ and 10 GPa, respectively. The total DOS near the $E_F$ mainly consists of Co-3$d$ and Zr-4$d$ orbitals. Figures 6(d) and 6(e) show the details of Co-3$d$ partial DOS at $P = 0$ and 10 GPa, respectively. The most of the partial Co-3$d$ DOS are located below the $E_F$ at both $P = 0$ and 10 GPa. However, the DOS for the Co-3$dz^2$ orbital presents above the $E_F$, which is an unoccupied state.

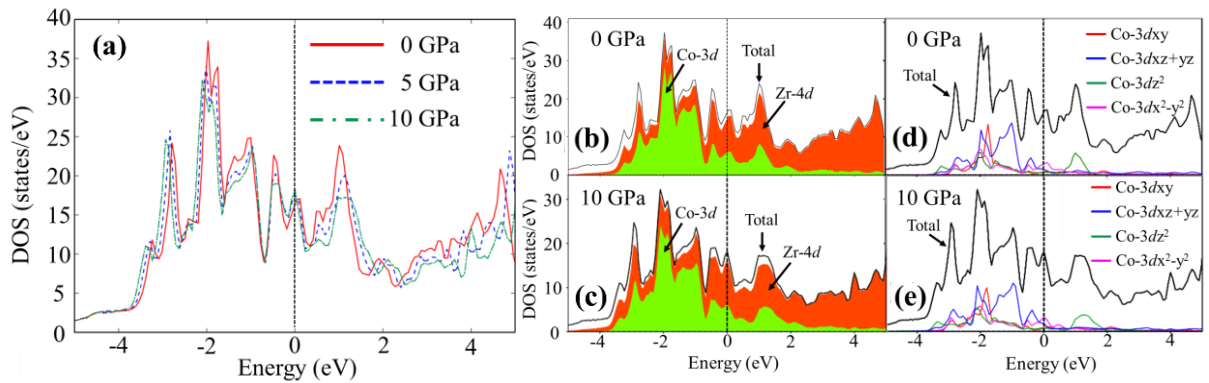

**Figure 6**. (a) Total Electrical density of states (DOS) for CoZr$_2$ at $P = 0$, 5, and 10 GPa. (b,c) Partial DOS for Co-3$d$ and Zr-4$d$ orbitals at $P = 0$ and 10 GPa, respectively. (d,e) Partial DOS for Co-3$d$xy, 3$d$xz+yz, 3$dz^2$, and 3$dx^2$-y$^2$ orbitals at $P = 0$ and 10 GPa, respectively.

Figure 7(a) shows the charge density isosurface of $CoZr_2$ at $P$ = 0, 3, 5, and 10 GPa. As pressure is applied, the charge density isosurface increases between Co-Co and Co-Zr bonds, resulting in a decrease in the bond lengths. The Co-Co and Co-Zr bond lengths are reversed at $P$ = 10 GPa (Figure 7(b)). Figure 7(c) shows the pressure dependence of bond lengths normalized by a value at ambient pressure. The normalized Co-Co bond length becomes shorter than the normalized Co-Zr bond length under pressures, suggesting that $CoZr_2$ flexibly contract along the $c$-axis as compared to the $a$-axis as shown in Figure 2(e). The change in the bonding state with increasing pressure would affect to the suppression of uniaxial PTE along the $a$-axis, which leads volumetric NTE.

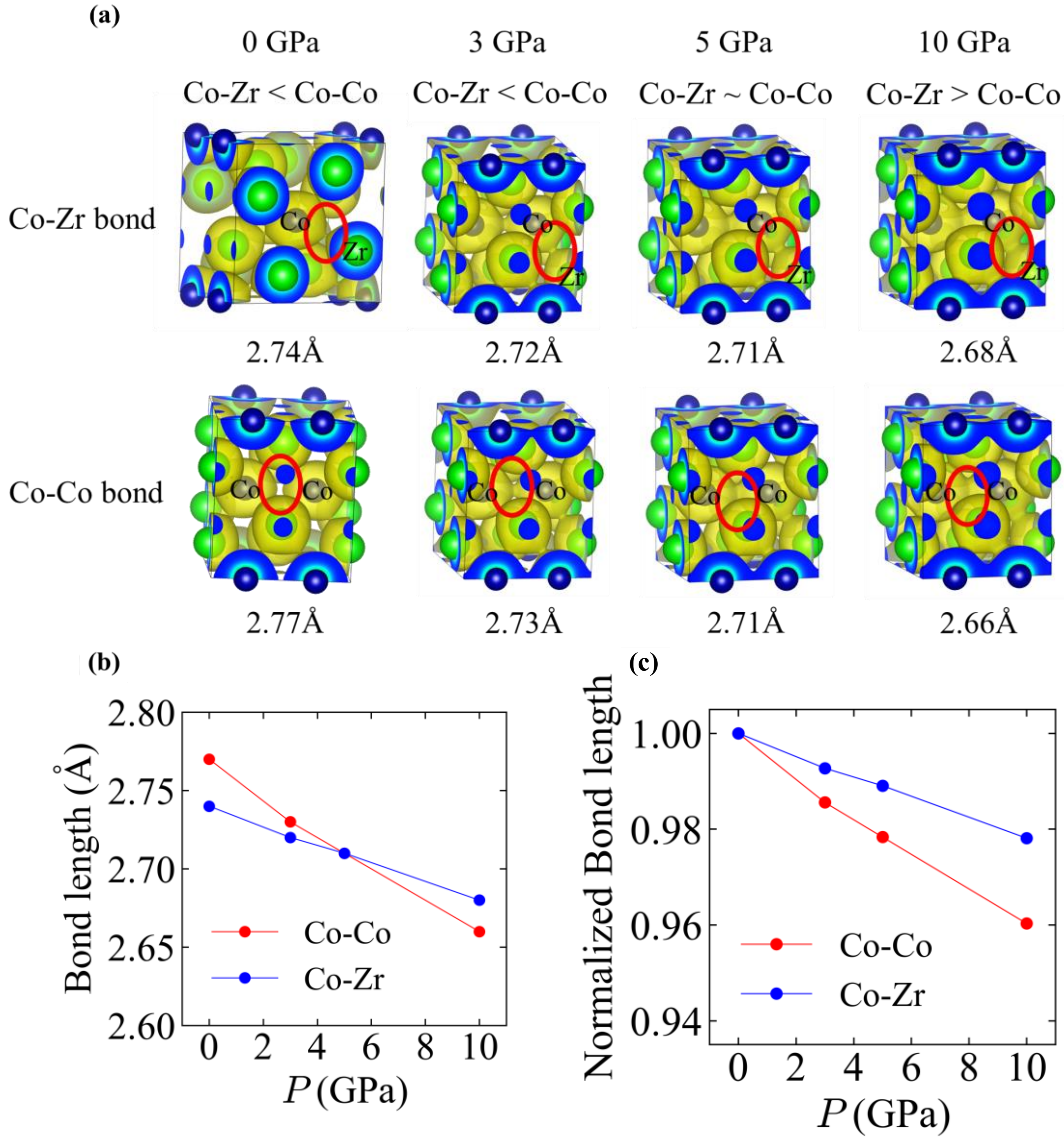

**Figure 7.** (a) Charge density isosurface of $CoZr_2$ at $P$ = 0, 3, 5, and 10 GPa. The bond length comparison between Co-Zr and Co-Co is shown at each applied pressure. (b) Pressure dependence of Co-Co and Co-Zr bond lengths. (c) Bond lengths normalized by a value at ambient pressure.

3. Conclusion

We measured temperature-dependent HP-SXRD patterns and the temperature dependence of $R$ for polycrystalline $CoZr_2$ under various pressures. From HP-SXRD results, we found that there was no crystal structural transition in the measured pressure and temperature ranges. The most important result of this study is the discovery of volumetric NTE in $CoZr_2$ induced by an application of high pressure. The uniaxial PTE along the $a$-axis was suppressed by pressure whereas the uniaxial NTE along the $c$-axis was not. As a consequence of the competition of uniaxial PTE and NTE, the volumetric NTE was achieved under high pressures in $CoZr_2$. The mechanisms of volumetric NTE would be related to the flexible crystal structure caused by the soft Co-Co bond, which has been proposed in a structural analogue $FeZr_2$, which also shows a giant uniaxial NTE along the $c$-axis under ambient pressure. From pressure dependence of $R$, we observed the dome-shaped pressure dependence of $T_c$ and an increase of $\Theta_D$ with pressure. The dome-shaped pressure dependence of $T_c$ would be caused by the competition between the pressure evolutions of $\Theta_D$ and $N(0)$. The increase of $\Theta_D$ against pressure could contribute to the suppression of uniaxial PTE along the $a$-axis. From electronic structure calculations, we found that the Co-Co and Co-Zr bond lengths were reversed at $P = 10$ GPa. The change in bonding state under pressures would be related to the emergence of unique axis thermal expansions of $CoZr_2$ under high pressures. Further investigation of phonon states potentially coupled with uniaxial NTE along the $c$-axis in $CoZr_2$ and its pressure dependence are needed to understand the mechanisms of volumetric NTE in $CoZr_2$ under high pressure. Through the systematic investigation on structural and physical properties of a $CoZr_2$ superconductor, we concluded that $CoZr_2$ exhibits the coexistence of superconductivity and volumetric NTE, which is possibly maintained in the temperature range lower than room temperature. The discovery will lead to the material exploration with volumetric NTE under ambient pressure in $TrZr_2$ and related superconducting material family.

4. Experimental Section/Methods

*Sample preparation*:

A polycrystalline sample of $CoZr_2$ was prepared using a Co rod (99.98 %, Nilaco) and Zr plates (99.2 %, Nilaco) by the Arc melting method. The sample chamber was filled with Ar gas after three times gas replacements. We synthesized the sample on a water-cooled Cu stage and turned it over several times when each melting step for homogenization.

*XRD and HP-SXRD measurements*:

The laboratory XRD patterns at ambient pressure were measured by the θ-2θ method with Cu-Kα radiation using a Miniflex-600 (RIGAKU) diffractometer equipped with a high-resolution semiconductor detector D/tex-Ultra. A BTS-500 attachment controlled the sample temperature. We measured HP-SXRD at the BL10XU beam line of SPring-8 with a wavelength of 0.413278 Å (Proposal No.: 2023A1254). The sample was loaded into a DAC with a pressure medium, He gas. The actual pressure was determined by the shift of the ruby R1 fluorescence line.[42] A band heater was mounted around the DAC to heat the sample. An K-type thermocouple was placed on the gasket to measure the temperature. Collected laboratory XRD and HP-SXRD patterns were refined by the Rietveld method using RIETAN-FP.[43] The images of crystal structure were depicted using VESTA.[44]

*Magnetic susceptibility measurement:*

The temperature and field dependences of magnetic susceptibility were measured using an MPMS3 (Quantum Design), a superconducting quantum interference device magnetometer. The temperature dependence was measured at $\mu_0 H = 1$ mT after both zero-field cooling (ZFC) and field cooling (FC) protocols. The field dependence was measured at $T = 1.8$ K.

*High-pressure electrical resistance measurement*:

The $T$ dependence of $R$ under various pressures were conducted using DAC with the boron-doped diamond micro-electrodes[45,46] in a PPMS (Quantum Design). The powders of cubic BN were filled in a hole around a metal gasket of SUS316 as a pressure medium. The generated pressures were determined from the shift of the ruby R1 fluorescence line[42] and the Raman peak from the diamond of a culet surface.[47]

*First principles calculation*:

First-principles calculations were performed using the VASP software package, employing the projector augmented wave method.[48-51] The Perdew-Burke-Ernzerhof exchange-correlation functional[52] was employed. K-point meshes of 11×11×12 and 22×24×24 were utilized for the internal coordinate optimization and the density of states calculations, respectively. Calculations were performed using the experimentally determined lattice constants. A plane wave cutoff energy of 350 eV was set. The partial density of states was visualized using PyProcar.[53]

Supporting Information

Supporting Information is available from the Wiley Online Library or from the author.

Acknowledgements

This work was partly supported by a Grant-in-Aid for Scientific Research (KAKENHI) (Proposal Nos. 21K18834, 23KK0088, and 23K13549), JST-ERATO (JPMJER2201), TMU Research Project for Emergent Future Society, and Tokyo Government-Advanced Research (H31-1).

# Supporting Information

**Pressure-induced volumetric negative thermal expansion in $CoZr_2$ superconductor**

*Yuto Watanabe*[A], *Hiroto Arima*[A], *Saori Kawaguchi-Imada*[B], *Hirokazu Kadobayashi*[B], *Kenta Oka*[B], *Hidetomo Usui*[C], *Ryo Matsumoto*[D], *Yoshihiko Takano*[D], *Takeshi Kawahata*[E], *Chizuru Kawashima*[E], *Hiroki Takahashi*[E], *Aichi Yamashita*[A], *Yoshikazu Mizuguchi*[A*]

A. Department of Physics, Tokyo Metropolitan University, 1-1, Minami–Osawa, Hachioji 192-0397, Japan.
B. Japan Synchrotron Radiation Research Institute (JASRI), 1-1-1 Koto, Sayo-cho, Sayo-gun Hyogo 679-5198, Japan.
C. Department of Applied Physics, Shimane University, Matsue, Shimane 690-8504, Japan
D. International Center for Materials Nanoarchitectonics (MANA), National Institute for Materials Science, Tsukuba, Ibaraki 305-0047, Japan.
E. Department of Physics, College of Humanities and Sciences, Nihon University, Setagaya, Tokyo 1568550, Japan.

E-mail: mizugu@tmu.ac.jp

## A. Thermal expansion of CoZr$_2$ at ambient pressure

We measured High-temperature X-ray diffraction (XRD) patterns at ambient pressure as shown in Figure S1. The usual volumetric positive thermal expansion was observed.

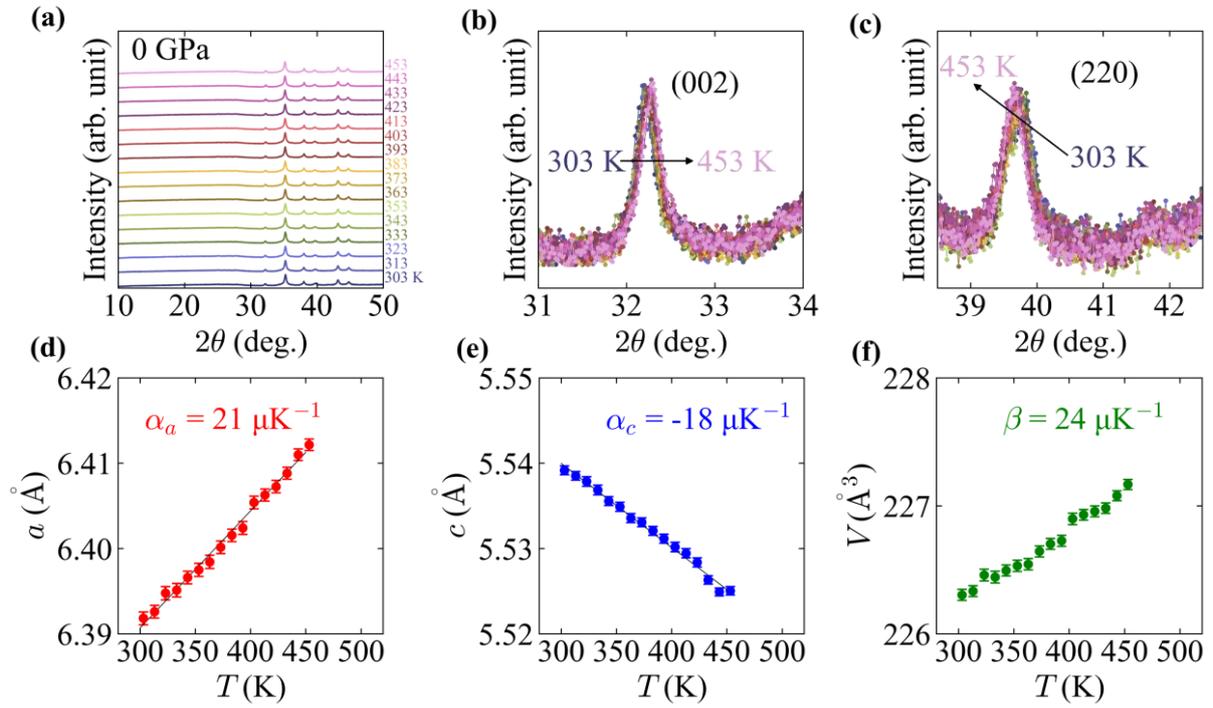

**Figure S1**. (a) High-temperature XRD patterns at $P = 0$ GPa (ambient pressure) for CoZr$_2$. (b,c) Shifts of 002 and 220 peaks due to heating. (d,e) Temperature dependence of lattice constants (d) $a$ and (e) $c$. The solid line is the fit to the linear line. (f) Temperature dependence of lattice volume $V$.

## B. Pressure dependence of lattice constants *a* and *c* of CoZr$_2$ at room temperature

Figure S2 (a) shows the pressure dependence of lattice constants *a* and *c* measured on another sample (different DAC from that shown in Fig. 2 in the main text) with a laboratory XRD. The normalized *a* and *c* by the ambient pressure are shown in Figure S2 (b). We observed the same trend of *a* and *c* against pressure, which is seen in Figure 2. A 4:1 mixture of methanol and ethanol is used for pressure transmitting medium to keep hydrostatic condition up to 10 GPa at room temperature. A Mo-Kα radiation is used to study larger *d*-space.

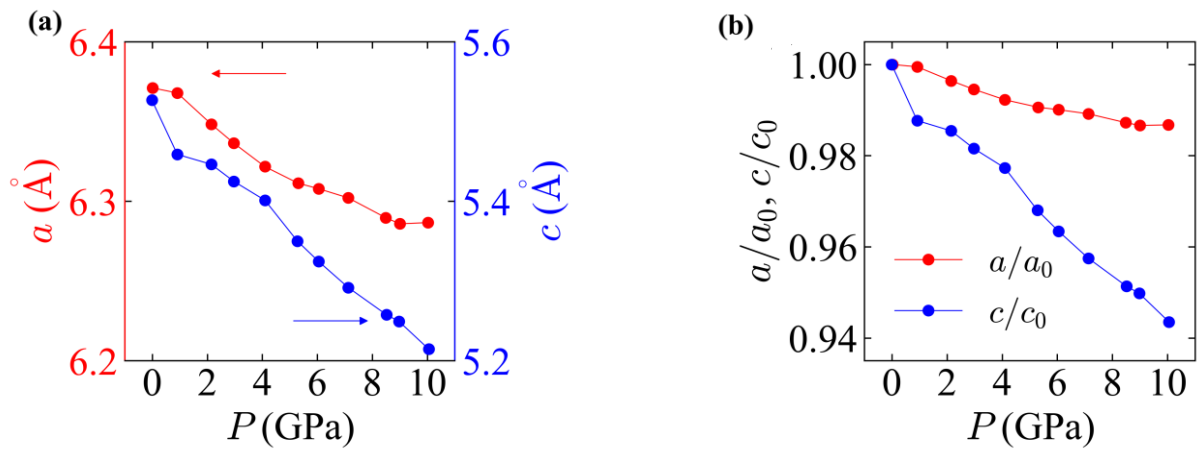

**Figure S2**. Pressure dependences of (a) lattice constants *a* and *c* and (b) lattice constants normalized by the ambient pressure. These data were collected on another sample (different DAC from that shown in Fig. 2 in the main text) with a laboratory XRD.

## C. High-pressure synchrotron X-ray diffraction patterns at various temperatures

Figure S3 shows the High-Pressure Synchrotron X-ray diffraction (HP-SXRD) patterns at various temperatures. There was no crystal structural transition in the pressure and temperature ranges examined in this study.

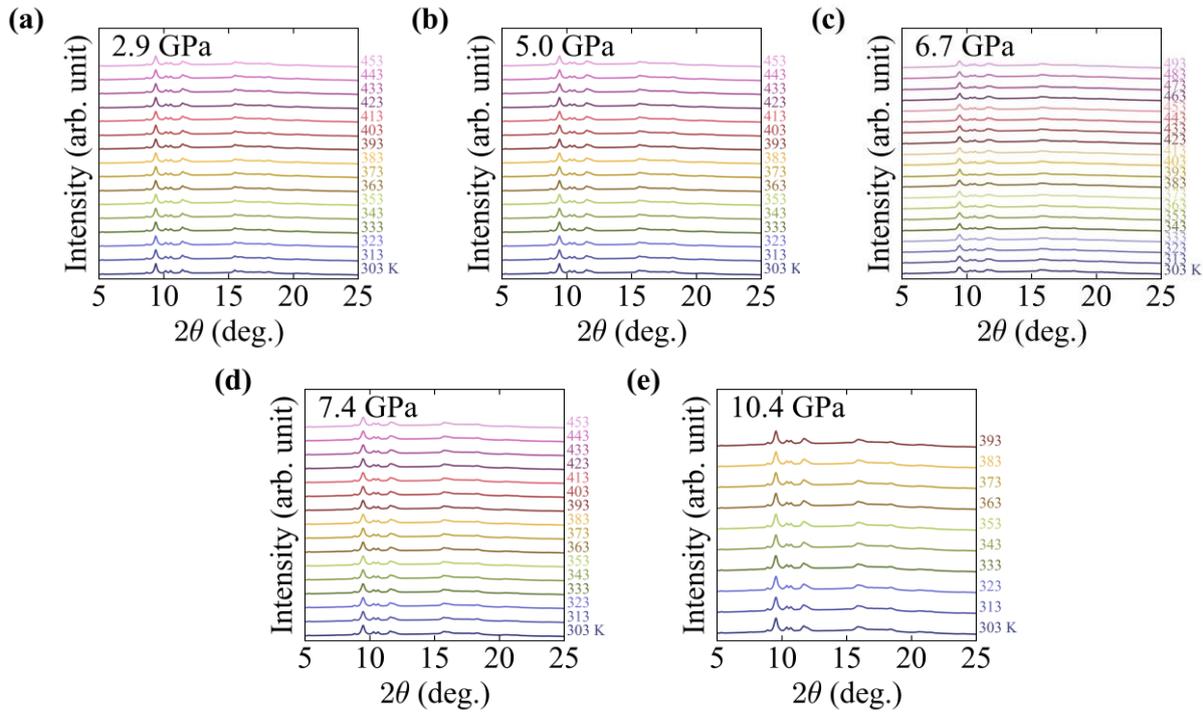

**Figure S3**. HP-SXRD patterns at (a) $P$ = 2.9 GPa, (b) $P$ = 5.0 GPa, (c) $P$ = 6.7 GPa, (d) $P$ = 7.4 GPa, and (e) $P$ = 10.4 GPa.

## D. Rietveld refinement for HP-SXRD patterns

Figure S4 shows the results of the Rietveld refinement of $CoZr_2$ at $P = 2.9$ GPa ($T = 303$ K, 453 K) and 10.4 GPa ($T = 303$ K, 403 K).

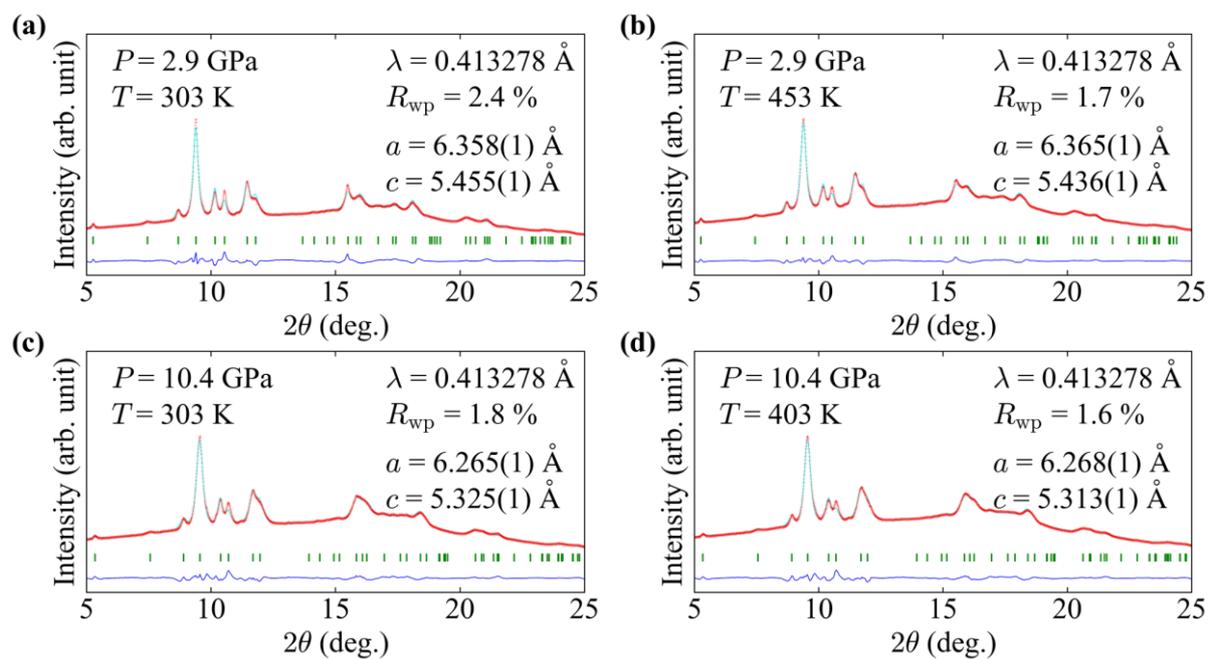

**Figure S4**. Rietveld refinement of HP-SXRD patterns at (a) $P = 2.9$ GPa, $T = 303$ K, (b) $P = 2.9$ GPa, $T = 453$ K, (c) $P = 10.4$ GPa, $T = 303$ K, and (d) $P = 10.4$ GPa, $T = 403$ K.

## E. Magnetic susceptibility of CoZr$_2$

We confirmed the bulk nature of superconductivity of CoZr$_2$ through magnetic susceptibility measurements as shown in Figure S4. The obtained superconducting transition temperature ($T_c$) was 5.7 K, which is consistent with the result of the electrical resistance measurement.

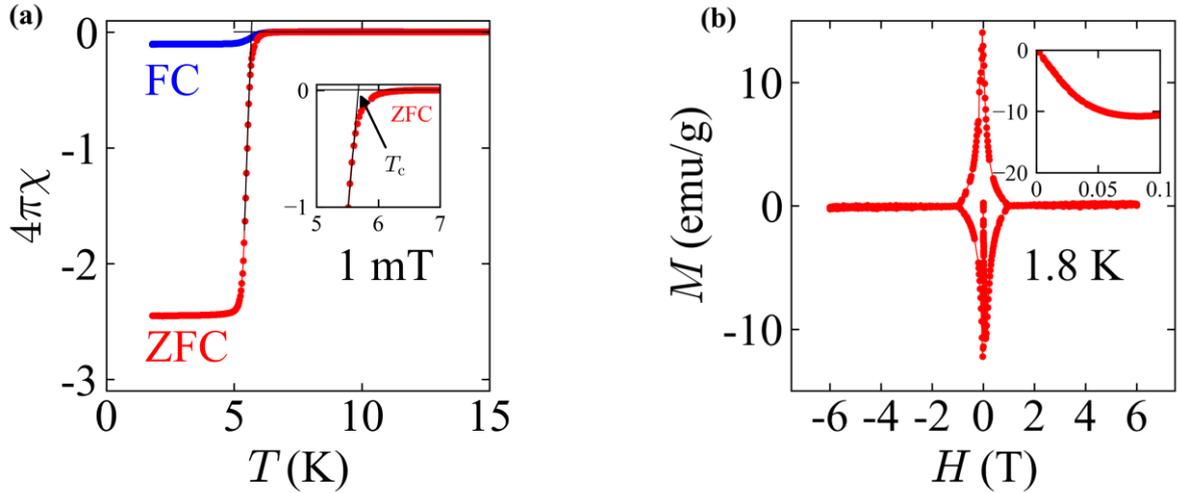

Figure S5. (a) Temperature dependence of magnetic susceptibility with zero-field cooling (ZFC) and field-cooling (FC) paths at $\mu_0 H = 1$ mT. (b) Field dependence of magnetic susceptibility at $T = 1.8$ K. The inset is the enlarged view in the Meissner-state regime.

## F. The Power-law relation in high-pressure electrical resistance

We fitted the electrical resistance ($R$) data in 10 K $< T <$ 30 K to the Power-law relation as the following equation.

$$R(T) = R_0 + AT^n \qquad (S1)$$

where $R_0$ is the residual resistance, $A$ is the numerical temperature-independent coefficient, and $n$ is a component depending on the carrier scattering mechanisms. The obtained component values were almost constant at $n = 3$ as shown in Figure S6 (b).

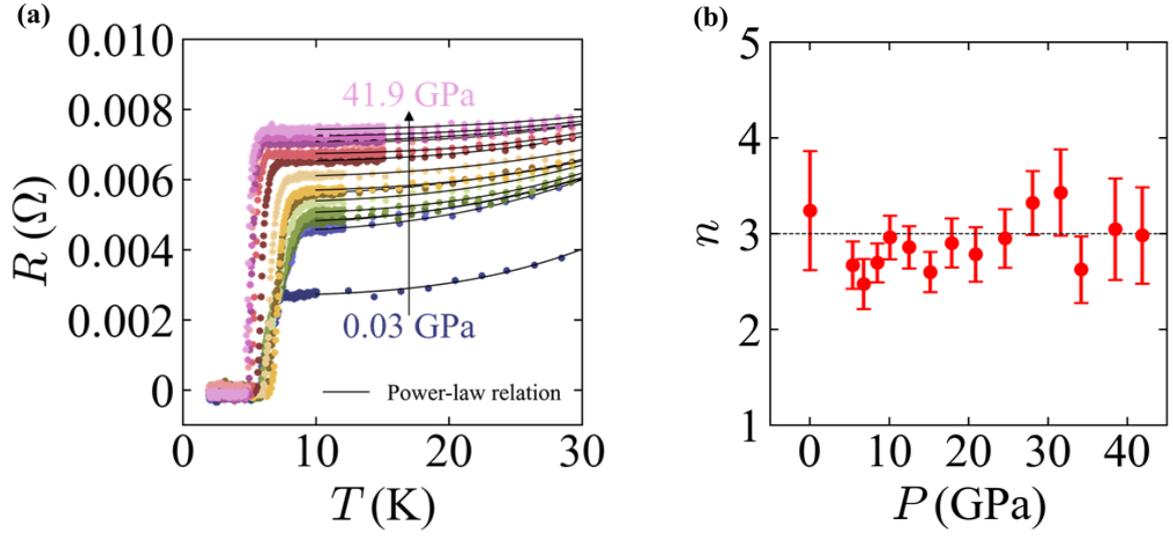

**Figure S6**. (a) Temperature dependence of $R$ near the superconducting transition. The solid line is fit to the Power-law relation (b) Pressure dependence of calculated exponent $n$.